# Measurements of group velocity of light in the lake Baikal water

Lubsandorzhiev B.K*., Pokhil P.G., Vasiliev R.V., Vyatchin Y.E.

*Institute for Nuclear Research of RAS*

*pr-t 60th Anniversary of October, 7A, 117312 Moscow, Russia.*

∗ *Corresponding author: postal address: pr-t 60$^{th}$ Anniversary of October, 7a, 117312 Moscow, Russia; phone: +7-095-1353161; fax: +7-095-1352268;*

*E-mail:* lubsand@pcbai10.inr.ruhep.ru

**Abstract**

The results of direct measurements of group velocity of light in the lake Baikal water at the depth of 1100 m are presented. The lake Baikal water dispersion has been measured at three wavelengths: 370 nm, 470 nm and 525 nm. The results are in a rather good agreement with theoretical predictions.

For the last few years the importance to distinguish group and phase velocity of light in high energy neutrino telescopes has been realized [1,2]. One can write well known equations cited in innumerable books, e.g. [3,4]. Phase velocity of light is defined as:

$$V_{ph} = \omega/k, \quad (1)$$

whereas group velocity is given by the following expression:

$$V_{gr} = \partial\omega/\partial k, \quad (2)$$

where $\omega$ and $k$ – wave frequency and wave vector of light correspondingly. Taking into account the trivial relations:

$$k = \omega n/c, \quad (3)$$

$$\omega = 2\pi c/\lambda, \quad (4)$$

where $c$ – velocity of light in vacuum, $n$ – refractive index of medium and $\lambda$ – wavelength of light; one can get expressions for phase and group velocities as follows:

$$V_{ph} = c/n, \quad (5)$$

$$V_{gr} = c/(n - \lambda \partial n/\partial \lambda). \quad (6)$$

Comparing the expressions (5) and (6) group refractive index $n_{gr}$ can be introduced by an analogy with phase refractive index $n$:

$$n_{gr} = n - \lambda \partial n/\partial \lambda, \quad (7)$$

and the expression (6) can be rewritten as:

$$V_{gr} = c/n_{gr}, \quad (8)$$

Thus phase and group velocities of light don't coincide in a medium with dispersion. Since the value of $\partial n/\partial \lambda$ is a negative for water in the wavelength range of 300÷600 nm the propagation time of Cherenkov light pulse in water in accordance with (5) - (8) will be larger in comparison with the expected time calculated basing on phase velocity.

During the last two expeditions at the lake Baikal we have carried out direct measurements of group velocity of light in the lake Baikal water at the depth 1100m just in the site of neutrino telescope NT-200 [5]. The measurements have been done with the experimental string developed to test new deep underwater techniques and detection methods for next generation neutrino telescopes at the lake Baikal. The scheme of the experimental string is shown in fig.1.

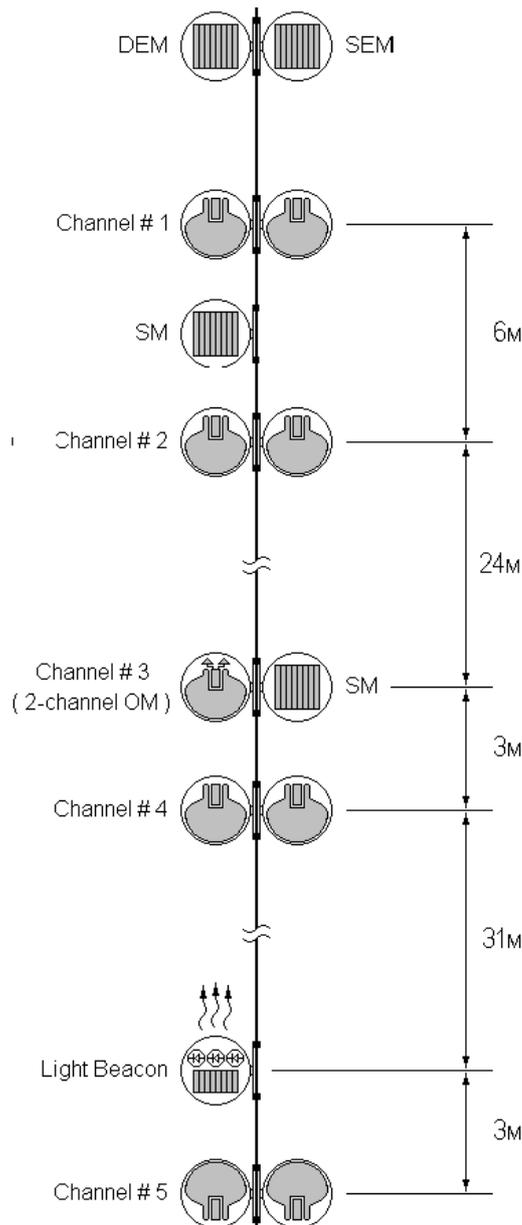

Figure 1. Scheme of experimental string. SM – front-end electronics module; SEM – string electronics module; DEM – detector electronics module.

This string consists of five optical channels – four standard optical channels of neutrino telescope NT-200 (channels #1, #2, #4, #5 in fig.1) and one pilot sample of new optical channel (channel #3). The optical channel of NT-200 is two optical modules [6] put in a pair and switched in coincidence. Each optical module is based on a hybrid phototube Quasar-370 [7]. The new experimental optical channel which is currently under development is based on a new two-channel version of Quasar-370 phototube [8]. The channel #5 was set to measure fully backscattered photons arrival time to estimate water scattering and absorption lengths and was not used in the dispersion measurements. The electronics system (Detector Electronics Module - DEM, String Electronics Module – SEM, front-end electronics module – SM) and the data acquisition (DAQ) system used in the measurements were analogous to the electronics and DAQ systems of NT-200 [5,6]. Light Beacons [9] developed for time and amplitude calibrations in NT-200 are used as light sources in these measurements. Light Beacon incorporates LEDs, LED drivers, pulse generator, controller, power supply etc encapsulated into glass pressure vessel. UV (NSHU590E), blue (NSPB300A) and green (NSPG300A) LEDs made by NICHIA CHEMICAL Ltd were used in Light Beacon. Emission spectra of these LEDs are presented in fig.2.

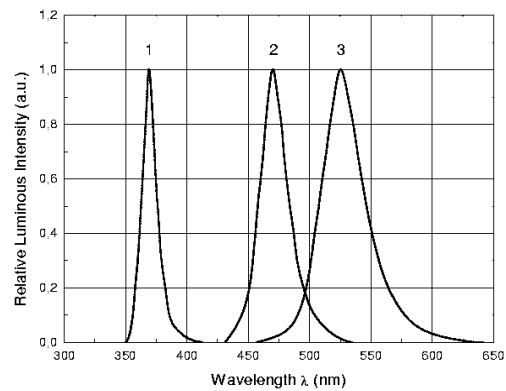

Figure 2. Emission spectra of NICHIA LEDs: 1 – NSHU590E; 2- NSPB300A; 3 – NSPG300A.

LED drivers [10] have been designed especially for the Baikal neutrino experiment. They exploit avalanche transistors FMMT415 produced by ZETEX and provide ~1ns width light pulses with up to $10^9$ photons per pulse.

We measured time intervals between trigger signals of LED drivers and time responses of optical channels with corrections to time delays in electronics and cables measured in the laboratory at the shore station of the lake Baikal neutrino experiment. The precision of the time measurements with described string was better than ±0.2 ns. The distances between light sources and optical channels were defined with precision of ~10 cm. The water dispersion is inferred from measurements of light pulses propagation times along the string and determination of group refractive index and group velocity of light and by taking into account the expressions (6)-(8).

The experimental results of $V_{gr}$ and $n_{gr}$ measurements for three wavelengths are collected in table 1.

Table 1.

| λ, nm | $V_{gr} *10^8$, m/s | $n_{gr} = n-\lambda dn/d\lambda$ |
|---|---|---|
| 370 ± 6 | 2.148 ± 0.010 | 1.396 ± 0.007 |
| 470 ± 11 | 2.193 ± 0.009 | 1.367 ± 0.006 |
| 520 ± 17 | 2.206 ± 0.009 | 1.359 ± 0.006 |

The propagation time of light pulses along the string for LED with λ = 470 nm versus distance between the Light Beacon and optical channels at the string is presented in fig.3.

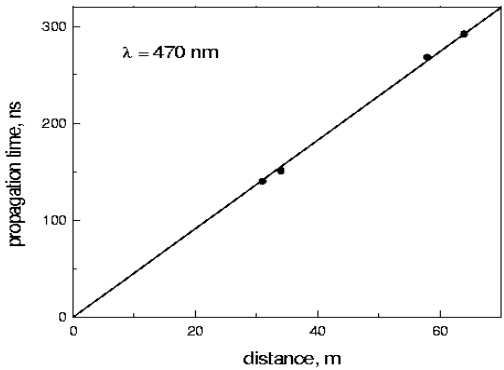

Figure 3. Dependence of propagation time of light pulses on distance along the string for λ = 470nm: • - experimental data; full line – approximation line.

The line is an approximation line of the experimental results (filled circles).

Fig.4 demonstrates the wavelength dependence of phase refractive index **n** for distilled water at a temperature of +20°C and a pressure of 1 atm (curve 1), the wavelength dependence of group refractive index $n_{gr}$ (curve 2) and the values of $n_{gr}$ (filled rectangles) experimentally measured for 370 nm, 470 nm and 525 nm wavelengths in the present work. Curve 1 was derived by

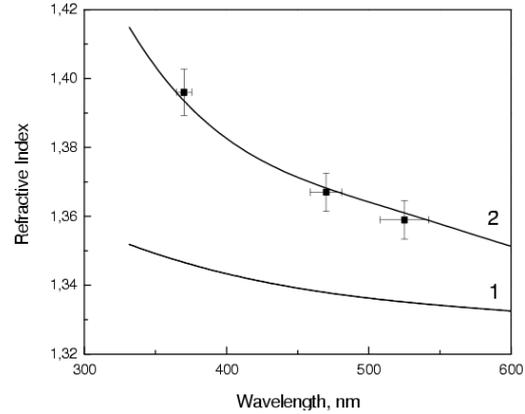

Figure 4. Wavelength dependences of **n** (curve 1) and $n_{gr}$ (curve 2) for distilled water. ■ - experimentally measured values of $n_{gr}$ for 370 nm, 470 nm and 525 nm.

approximating polynomially extensive experimental data of **n** taken from [11]:

$$n(\lambda) = 1.9672 - 0.59931 \cdot 10^{-2}\lambda + \\ + 0.24515 \cdot 10^{-4}\lambda^2 - 0.54076 \cdot 10^{-7}\lambda^3 + \\ + 0.66727 \cdot 10^{-10}\lambda^4 - 0.43317 \cdot 10^{-13}\lambda^5 + \\ + 0.11502 \cdot 10^{-16}\lambda^6, \qquad (9)$$

here λ in nm. The curve for $n_{gr}$ was calculated in terms of curve 1 with corrections to the neutrino telescope NT-200 conditions: a temperature of +3.4°C and a hydrostatic pressure of 110 atm corresponding to a depth of 1100 m. One can see a rather good agreement between the experimental results and the calculated curve. We would like to note here that the calculated curves corresponds to distilled water but it was pointed out in [1] that dispersion curves for the lake Baikal water in the waveband 300 nm ≤ λ ≤ 600 nm does not differ much from distilled water curves because the lake Baikal deep water has no intense absorption lines in this waveband [12].

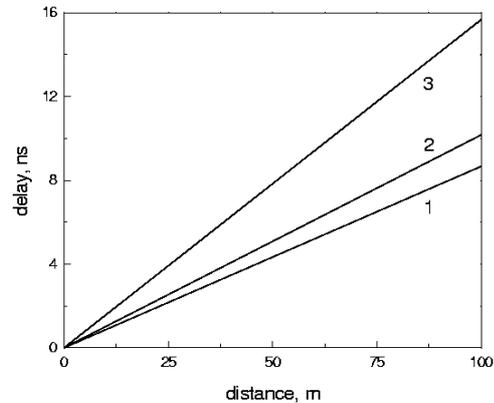

Figure 5. Dependence of time delay due to phase and group velocities disparity on distance for different wavelengths: 1 – 370 nm; 2 – 470 nm; 3 – 525.

The dependence of the light pulses time delay on the distance between the Light Beacon and optical channels due to phase and group velocities disparity for different wavelengths are shown in fig.5. The time delay between light pulses with $\lambda = 370$ nm and $\lambda = 525$ nm at a distance base of 100 m is seen to be equal to ~8 ns.

For present and future large scale water Cherenkov arrays like NT-200 [5], NT-200+[13] and Giant Volume Detector [14] in the lake Baikal, ANTARES [15], NESTOR[16] and NEMO [17] in the Mediterranean Sea with substantial spacing between strings and aiming to detect distant energetic showers due to high energy neutrino interactions it is of crucial importance to know the deep water dispersion for an adequate understanding of experimental data.

The authors would like to express gratitude to all their colleagues from Baikal Collaboration for help in these measurements, to Prof. E. Lorenz and Dr. Th.Schweizer for providing with UV LEDs, to Dr. L.A.Kuzmichev for introducing us with approximation (9). We are indebted to Dr. V.Ch.Lubsandorzhieva for careful reading of the paper and many invaluable remarks and advises. This research was supported in part by Russian Fund of Basic Research (grants № 02-02-17365, 02-02-31005, 02-02-17031, 00-15-96794)

**References.**

[1] L.A.Kuzmichev, Nucl. Instr. and Meth. A 482, (2002) 304.
[2] P.B.Price, K.Woschnagg, Astropart. Phys. 15 (2001) 97.
[3] A.Sommerfeld, OPTIK, Wiesbaden, 1950.
[4] L.D.Landau, E.M.Lifshiz, Electrodynamics of Continuous Media, Nauka, Moscow, 1991.
[5] I.A.Belolaptikov et al., Astropart. Phys. 7 (1997) 263.
[6] R.I.Bagduev et al., Nucl. Instr. and Meth., A 420, (1999) 138.
[7] R.I.Bagduev et al., Proceedings if the International Conference on "Trends in Astropart. Phys.", 1994, p. 132.
[8] B.K.Lubsandorzhiev et al., Proceedings of the 27[th] ICRC, V3, p. 1294.
[9] B.K.Lubsandorzhiev et al., Proceedings of the 27[th] ICRC, V3, p.1291.
[10] R.V.Vasiliev et al., Instr. and Experim. Tech. 43 (2000) No. 4, p.570.
[11] I.S.Grigoriev, E.Z.Meilikhov (Eds.), Physical Quantities, Handbook, Energoatomizdat, Moscow, 1991, p. 790.
[12] B.A.Tarashansky. PhD Thesis. INR Moscow. 1999.
[13] V.A.Balkanov et al., Proceedings of the 27[th] ICRC, V3, p.1096.
[14] V.A.Balkanov et al., Proceedings of the 20[th] International Conference on Neutrino Physics and Astrophysics, Neutrino 2002, Munich, Germany, May 25-31, 2002
[15] P.Amram et al., Proceedings of the 27[th] ICRC, Hamburg, Germany, August 2001, V.3, P.1233.
[16] E.Anastosis et al., Proceedings of the 25[th] ICRC, Durban, South Africa, 1997, V.7, P.49.
[17] C.De Marzo et al., Proceedings of the 6[th] Intern. Workshop on Topics in Astroparticle and Underground Physics. Paris, France, 6-10 September 1999, P.433.